\documentstyle[aps,prb,tighten,preprint]{revtex}
\begin{document}
\draft
\title{Quantum railroads and directed localization
at the juncture of quantum Hall systems}
\author{Shinji Nonoyama$^{1,2}$
and George Kirczenow$^{1}$}
\address{$^{1}$Department of Physics, Simon Fraser
University, Burnaby, B.C., Canada
V5A 1S6\\$^{2}$Faculty of Education, Yamagata
University, Yamagata 990-8560, Japan}
\maketitle
\begin{abstract}

The integer quantum Hall effect (QHE) and
one-dimensional Anderson localization (AL) are
limiting special cases of a more general
phenomenon, directed localization (DL), predicted
to occur in disordered one-dimensional wave guides
called ``quantum railroads" (QRR). Here we explain
the surprising results of recent  measurements by
Kang {\em et al.} [Nature {\bf 403}, 59 (2000)] of
electron transfer between edges of two-dimensional
electron systems  and identify experimental
evidence of QRR's in the general, but until now
entirely theoretical, DL regime that unifies the
QHE and AL. We propose direct  experimental tests
of our theory.

\end{abstract}


\section{Introduction}

It was discovered by von Klitzing, Dorda and
Pepper\cite{vonKlitzing} that the Hall conductance of a
two-dimensional electron gas (2DEG) in a magnetic field is
quantized in integer multiples of the universal quantum
$e^2/h$. This remarkable phenomenon was explained by
Laughlin\cite{Laughlin} using a gauge invariance argument.
Subsequently, however, Streda, Kucera and
MacDonald,~\cite{Streda} Jain and Kivelson\cite{Jain} and
B{\"u}ttiker\cite{Buettiker} proposed an alternate point of
view in which the integer quantum Hall effect (QHE) is
explained on the basis of the Landauer theory\cite{Landauer}
of one-dimensional (1D) transport, within the framework of
magnetic edge states introduced by Halperin.~\cite{Halperin}
These states which derive from the quantized Landau levels
of the 2DEG in a strong magnetic field follow the edges of
the sample and are the 1D transport channels of these
theories. The electrons in these edge channels travel in the
{\em same} direction and therefore cannot be backscattered.
Thus as was elucidated by B{\"u}ttiker,~\cite{Buettiker} they
are immune from the effects of Anderson localization that
normally inhibits propagation of waves or quantum particles
through 1D wave-guides in the presence of
disorder\cite{Anderson,Mott,KM}: They travel along the edge
of a disordered sample over macroscopic distances without
resistance or dissipation of energy.

Although the QHE and 1D Anderson localization (AL) are
mutually antithetical in this fundamental way, it has been
shown \cite{Barnes} that they are both special cases of a
more general phenomenon, namely, {\em directed} localization
(DL). This is a property of disordered 1D wave-guides called
``quantum railroads"(QRR) that support {\em arbitrary}
numbers of channels carrying electrons in {\em opposite}
directions.~\cite{Barnes,periodic} If $L$ channels carry
electrons from New York to Los Angeles and $M$ in the
opposite direction, it is predicted\cite{Barnes} that for $L
\le M$, none of the electrons leaving New York will reach
Los Angeles: After multiple scattering events they all
return to New York. However of $M$ electrons leaving Los
Angeles on average $M-L$ will reach New York while $L$
return to Los Angeles. Thus in such systems the physics of
localization acquires a directionality: All electrons
traveling in the minority direction behave as if they are
localized while some of those traveling in the majority
direction are transmitted through the macroscopic disordered
system. AL and the QHE correspond to the special cases $M =
L$ and $L = 0$, respectively. These important effects have
received much attention.~\cite{KM,PG,SP} However, there have
been no reports of experimental realizations of the general
DL regime ($M$$>$$L$$>$$0$) that should unify
them.~\cite{Moon} Recent measurements by Kang {\em et
al.}\cite{Kang} of electron transfer between the edges of two
2DEG's revealed a richness of unexpected and puzzling phenomena.
In this article we use computer simulations to identify the
physics behind these observations. We demonstrate that a
consistent explanation of the experiment is possible only if the
barrier between the 2DEG's is surrounded by a potential well
that supports QRR's of edge channels exhibiting DL in the general
$M$$>$$L$$>$$0$ regime. Interplay between DL and electron
transfer is  reflected directly in the data. Unlike previous
theoretical work, the present theory accounts for {\em
all} of the features of the data. It admits simple and
direct experimental tests.

In Section \ref{puzzles} we outline some key puzzles
posed by the experiment of Kang {\em et al.}\cite{Kang}
and present a critique of an important assumption that
has been made in previous theoretical attempts to
explain the data. In Section \ref{disorder} we examine
the effects of disorder on electron transfer through the
barrier. In Section \ref{charge} we present
self-consistent Hartree calculations of the edge channel
energies in the presence of disorder and arrive at a
model  in which the barrier between the 2DES's is
surrounded by a strong potential well. In Sections
\ref{bias} and \ref{breakdown} we solve  this model in
the regime of finite applied bias voltages and show how
the physics of directed localization in QRR's in concert
with the breakdown of the quantum Hall effect can
account for all of the phenomena observed by Kang {\em
et al.}\cite{Kang} under applied bias. Our concluding
remarks are presented in Section \ref{conc}.

\section{Puzzles posed by the experiment}
\label{puzzles}

The energies of the 2DEG edge states near the barrier
that separates the 2DEG's are shown in Fig.\ref{fig1}a
for the simplest model\cite{Kang} of the experimental
system in which disorder and interactions between
electrons are neglected and no bias is applied between
the 2DEG's. Curves with positive and negative slope
correspond to edge channels on opposite sides of the
barrier, and electrons in these channels travel in
opposite directions as shown in the left inset of Fig.1a.
Due to conservation of energy and momentum only
electrons at crossings of the curves in Fig.1a can
transfer through the barrier. At the crossings
themselves there are small energy gaps shown in the
right inset of Fig.\ref{fig1}a. There are no extended
states at energies in the gaps. Therefore an electron in
a gap cannot travel far along the barrier. Since it also
cannot reverse its course it {\em must} pass through the
barrier. Thus for a long, high barrier electrons with
energies in the gaps should be transmitted perfectly
through the barrier while electrons at other energies
should hardly be transmitted at all.  As was noted in
Ref. \onlinecite{Kang} the predictions of this simple model
disagree with the data: The differential conductance
peaks at zero bias (that signal electron transfer
through the barrier) persist over ranges of magnetic
field far larger than expected from the small widths of
the energy gaps. Also the peaks occur at magnetic fields
larger than predicted by factors of 2-4. Kang {\em et
al.} conjectured\cite{Kang} that these discrepancies may
be due to disorder, unexpectedly strong spin
polarization or a potential well near the barrier.

Pioneering theoretical  studies\cite{TP,MG,LY,KS,KF} based on more
detailed models have addressed the roles of transport and
electron-electron interactions in this system. In these
theories,\cite{TP,MG,LY,KS,KF} in order to account
for the observed positions\cite{Kang} of the zero-bias conductance
peaks,   the 2DEG's were assumed to be {\em fully} spin-polarized,
{\em at least} in the range of Landau level filling
factors $\nu$ between 1.1 and 1.5 where the first
zero-bias conductance peak was observed. However this
assumption is difficult to reconcile with the absence of
features due to spin in the data\cite{Kang} and also with
the presence in the data of a prominent feature that will
be discussed in Section
\ref{breakdown}. Physically the assumption means that
the {\em second} 2DEG Landau level must be partly occupied
by electrons having one spin orientation before electrons
with the opposite spin orientation begin to occupy their
{\em lowest} Landau level. A necessary ({\em although not
sufficient}) condition for this to occur is that the
electron spin splitting energy must be
larger than the Landau level splitting, i.e.,
$g \mu_B B / \hbar \omega_c > 1$.  But this condition is
{\em not} satisfied by the values of the 2DEG g-factors that
have been measured
experimentally\cite{Nicholas,Goldberg,Dolgopolov,Wiegers}:
The measured
values\cite{Nicholas,Goldberg,Dolgopolov,Wiegers}  of the
exchange-enhanced\cite{AU} Land\'{e}  factor
$g$ in 2DEG's in GaAs have ranged up
to  a maximum value
of approximately 6 that corresponds to $g\mu_B B/\hbar
\omega_c = 0.2$. Larger $g$-factors have been reported in quasi-{\em
one}-dimensional quantum wires but the the {\em largest} measured
$g$ values even in those systems have not been large enough to be
consistent with a fully  spin-polarised system with more than one
occupied Landau level.~\cite{wire} Thus it seems unlikely that the
2DEG's of Kang {\em et al.}\cite{Kang} were fully spin polarized when
more than one Landau level was occupied. Therefore, in this article we
explore an alternate possibility that is consistent with the
experimentally established properties of the
$g$-factors of 2DEG's that were summarized
above.\cite{Nicholas,Goldberg,Dolgopolov,Wiegers} Namely, we assume that
the spin splitting in the experiment of Kang {\em et al.}~\cite{Kang}
was considerably smaller than
$\hbar
\omega_c$ and also  smaller than the Landau level
broadening  (that is due to disorder) so that the spin
splitting can be neglected as a first approximation. We will show in
Section
\ref{charge} that if this assumption applies then the zero bias data of of
Kang {\em et al.}~\cite{Kang} can be explained only if the barrier in
their device was strongly charged. However, in that case we are able to
explain not only the observations of of Kang {\em et al.} at zero
bias,~\cite{Kang} but also all of their results at finite
bias~\cite{Kang} including those that have not been accounted for by
any previous model.

On the basis of their theoretical analysis Mitra and
Girvin\cite{MG} have argued that in order to account for
the persistence of the experimentally
observed\cite{Kang}   zero-bias conductance peaks it is
necessary to consider the effects of disorder on electron
transfer through the barrier. However, they did not address this issue
quantitatively. We shall do this in the next Section.

\section{Electron transfer through the barrier in
the presence and absence of disorder}
\label{disorder}

Fig.\ref{fig2} shows results of our computer simulations
of electron transfer through the barrier obtained using
a recursive Green's function technique.~\cite{NO} The
solid curve in Fig.\ref{fig2}a is the transfer
probability $T$ in absence of disorder and e-e
interactions. It confirms the paradoxical predictions of
the simple model that were outlined at the beginning of
Section \ref{puzzles}:  At energies in the gap (shown in
the inset, Fig.\ref{fig1}a), electrons are transmitted
perfectly through the barrier and $T = 1$. Elsewhere $T$
is near zero except for oscillations due to quantum
interference near the main peak; see the inset
Fig.\ref{fig2}a. The period of the oscillations is two
orders of magnitude smaller here than in Ref.\onlinecite{TP}
due to the much longer barrier in our simulations.
Fig.\ref{fig2}b shows $T$ for the same system but with
an additional smooth random potential that models the
electron Coulomb interaction with donor ions in the
doped layer adjacent to the 2DEG's. Comparing
Fig.\ref{fig2}b and \ref{fig2}a shows that even a weak,
smoothly varying disordered potential alters the electron
transfer mechanism dramatically: $T$ now takes the form
of a dense array of extremely narrow resonances
associated with electron states in the random potential.
The dashed curve in Fig.\ref{fig2}a shows the combined
effect of finite temperature and disorder on the
transfer for the same potential as in Fig.\ref{fig2}b:
The transmission peak is now sufficiently broad to
explain the observed persistence of the conductance
peaks, and also the absence of features due to spin in
the data {\em if the spin-splitting is small as in our
theory}. However it has not shifted significantly from
its position in the absence of disorder. Our simulations
for other types of disorder and barrier profiles
confirmed that the positions of thermally broadened
transmission peaks coincide with the energy gaps of the
electron edge channels at the barrier, {\em regardless
of the nature of the disorder}. Thus the explanation of
the observed locations of the conductance
peaks\cite{Kang} must involve {\em other} properties of
the 2DEG and barrier that control the energies at which
the gaps occur.

\section{Electron-Electron Interactions and Charging at
the Barrier}
\label{charge}

To explore this further we performed self-consistent
Hartree calculations of the edge channel energies,
treating the effects of disorder in a mean-field
approximation: We assumed that disorder broadens the
density of states of each Landau level into a Gaussian
distribution whose center tracks the position-dependence
of the Hartree  potential.~\cite{explain_broadening} This
resulted in a downward shift of the edge channel
energies and gaps relative to their positions in
Fig.\ref{fig1}a due to the electron density in the
barrier being depleted relative to the 2DEG outside the
barrier and this depletion resulting in an electrostatic
well around the barrier.  However, the size of the shift
was insufficient to explain the observed\cite{Kang}
positions of the zero bias conductance peaks.

To characterize the size of the discrepancy
physically, we repeated our calculation assuming an
{\em additional} positive charge  density $\rho$ to
be present inside the barrier such as might be
introduced by doping the barrier in the plane where
it was cleaved during fabrication of the sample.
This yielded a downward shift of the energy gaps
sufficient to explain the observed
positions\cite{Kang} of the conductance peaks (for
{\em spin-unpolarized} 2DEG's) for
$\rho=11\times10^{11} e$ cm$^{-2}$. This value of
$\rho$ is so much (5.5$\times$) larger than the
charge density of the 2DEG that it is unreasonable
to expect {\em any} electronic many-body effect
(i.e., correction to Hartree theory) to result in
an effective potential well (whether spin-dependent
or not) that could mimic the electrostatic well
associated with this large positive charge that is
required to account for the experimental data. We
believe that this rules out {\em all} potential
explanations of the observed positions of the
conductance peaks that rely primarily on many-body
corrections, including any explanation that
involves strongly spin-polarized electron systems.
Since the sample was made using a new, still
incompletely understood process we consider the
possibility of a {\em materials-related} potential
well that we have modeled by the above charge
density $\rho$ to be reasonable. This potential well is
an important part of our model of the device of
Kang {\em et al.}\cite{Kang} that we solve below.
(Recall that previous theories~\cite{TP,MG,LY,KS,KF} relied
instead on the assumption that the 2DEG's are fully spin
polarized in a range of values of
$\nu >1$ to explain the observed
positions~\cite{Kang} of the conductance peaks at
zero bias. That
assumption, as was explained in Section
\ref{puzzles}, is not consistent with the experimental
literature on spin polarization in 2DEG's and has other
drawbacks.) We note that the possibility  that
some inadvertent doping may have occurred at the cleaved
interaface  is consistent with the measured mobility ($10^5$
cm$^2$ V$^{-1}$ s$^{-1}$) of the 2DEG's\cite{Kang} that is
unexpectedly low for a state of the art GaAs
device today. The disordered potential
associated with such doping would contribute to the
Landau level broadening discussed in Section
\ref{disorder} to which we and others\cite{MG} have
attributed the experimentally  observed\cite{Kang}
persistence of the differential conductance peaks
at zero bias. Furthermore disorder near the barrier plays a key role in
the  physics of directed localization\cite{Barnes}
that as we show in Section \ref{bias} is able to
account for the fact\cite{Kang} that tunneling
between some pairs of Landau levels is observed
experimentally while for other pairs it is not
observed. This is an important feature of the
data\cite{Kang} that to date has not been explained
in any other way.

\section{Electronic Structure and the Physics of
Directed  Localization at Finite Applied Bias}
\label{bias}

The theories proposed to date\cite{TP,MG,LY,KS,KF}
have either not treated the case where a finite
bias voltage was applied across the barrier in the
experiment of Kang {\em et al.}~\cite{Kang} at all
or have yielded qualitative inconsistencies  with
the data~\cite{Kang} in that regime. Here we shall
apply the  model with the charged barrier proposed
in Section \ref{charge} to the case of finite
applied bias voltages. We shall show that this
model is able to account for all of the features of
the data.~\cite{Kang}

Examples of our self-consistently calculated edge
channel  Hartree energies under applied bias are
shown in Fig.\ref{fig1}b,c. The bias $V$ results in
a difference $eV$ between the Fermi levels $E(S)$
and $E(D)$ on the source and drain side of the
barrier. The energy of each edge channel now
exhibits a {\em minimum} due to  the potential well
near the barrier. Since the electron velocity is
$v=dE/d(\hbar k)$ this means that edge channels can
carry electrons in {\em opposite} directions along
the {\em same} side of the barrier. Thus QRR's in
the {\em general}  ($M$$>$$L$$>$$0$) DL regime are
realized. Two examples are the QRR's near $E(S)$ on
the source side in  Fig.\ref{fig1}b and c. In both
cases there are 2 forward moving channels (the $N=0$
and 1 channels at $w,x,u$ and
$f$) and 1 backward channel (the $N=1$ channels at
$z$ and
$b$). Thus for an impenetrable barrier, DL theory
would predict\cite{Barnes} half of the electrons
entering the QRR in the forward direction to be
reflected and half to be transmitted through the
QRR. However, at edge channel crossings transfer
through the barrier competes with the forward and
back-scattering alternatives of DL. As will be
explained below, this competition results in {\em
selective} suppression of the transfer at certain
edge channel crossings, in agreement with the
conductance data.~\cite{Kang} For example, transfer
is suppressed at crossing
$x$ (denoted 1,0 in Fig.\ref{fig3}) but not at
$u$ (0,1 in Fig.\ref{fig3}). This difference can be
understood physically in terms of the unitarity of
the scattering matrix that underpins the
predictions of DL theory: Electrons enter the QRR
in the two forward channels. If the barrier is
impenetrable, then after multiple  scattering
events these electrons fill a single effective
forward channel and a backward channel. However the
$N=1$ forward edge states near $f$ and $x$ and the
backward edge states near $b$ and $z$ have low
velocities and are thus strongly mixed and
localized by disorder. Therefore the effective
forward channel of DL theory consists, for the most
part, of the $N=0$ edge state at $u$ or $w$. Now
consider a barrier that is {\em not} impenetrable.
Then it is crucial whether an energy gap opens
\underline{in the $N=0$ edge channel} at
$E(S)$, blocking propagation through the forward DL
channel.  This happens in Fig.\ref{fig1}c (the gap
is at $u$) so that electrons in the forward DL
channel are blocked and must transfer through the
barrier. But in Fig.\ref{fig1}b the gap is at
$x$ so the forward DL channel is still open (at
$w$) and forward propagation dominates over
transfer through the barrier which is only weakly
transmitting. Thus transfer at crossing
$x$ is suppressed. The absence of the associated
conductance peak in the data of Kang {\em et
al.}\cite{Kang} is evidence of a QRR in the general
($M$$>$$L$$>$$0$) DL regime: It is inconsistent
with models in which all edge states on the same
side of the barrier travel in the same direction
because for such models {\em unitarity} requires
that transfer through the barrier {\em not} be
suppressed at {\em any} edge channel crossing. This
is why the absence of this peak in the
data\cite{Kang} could not be explained by previous
theoretical work.~\cite{TP}

Backscattering also occurs at the crossing of the
two $N=0$ channels in Figs.\ref{fig1}b and c.
However this crossing is well below the source 2DEG
Fermi level. Thus, if the bias is somewhat larger
than in Fig.\ref{fig1}b so that the crossing is
above
$E(D)$, electrons from occupied states at higher
energies {\em along the entire length of the source
side of the barrier} can decay to states at this
crossing and then transfer through the barrier.
Thus we do not expect the conductance peak
associated with this crossing to be suppressed by
localization effects.

The locations of the conductance maxima predicted
by our calculations of the edge channel crossings
are compared with the observed positions\cite{Kang}
of the conductance peaks in Fig.\ref{fig3}. Blue
(red) curves indicate edge channel crossings at
which transfer is (not) expected to be suppressed
according to the above considerations. There is an
obvious one-to-one correspondence between the red
curves and the loci of experimental conductance
maxima and good quantitative agreement at Landau
level fillings $>$1.14 for positive and small
negative bias. However the red curves do not follow
the bell-shaped structure at lower Landau level
fillings or exhibit the asymmetry seen
experimentally between positive and negative bias.
In Section
\ref{breakdown} we explain these  deviations as
manifestations of the breakdown of the quantum Hall
effect (QHE).

\section{Breakdown of the quantum Hall effect}
\label{breakdown} The bias voltage in our
self-consistent  calculations is evaluated between
points in the vicinity of the barrier.  In the quantum
Hall regime the 2DEG's are perfect conductors.
Therefore the measured bias is equal to the bias
voltage in our theory. When the QHE breaks down, the
2DEG's become resistive and the measured bias acquires a
contribution due to potential differences within the
2DEG's {\em in addition} to the potential drop across
the barrier; thus the magnitude of the experimentally
measured bias voltage should exceed  the theoretical
value. If the magnetic spin-splitting is {\em small}
(as in our theory), the QHE breaks down at zero bias
when the Landau level filling falls below a value
somewhat larger than 1 that depends on the 2DEG
mobility.~\cite{PG,SP} Thus we explain the bell-shaped
structure in the experimental data below the Landau
level filling of 1.14 where the magnitude of the
experimental bias voltage at the conductance peak
abruptly begins to exceed  the theoretical one as a
manifestation of the breakdown of the QHE. We note that
no explanation of the bell-shaped structure was offered
by previous theories. The present explanation is not
compatible with their assumption that the 2DEG is fully
spin-polarized since in that case the QHE would break
down at a quite different value of the Landau level
filling. The value of the Landau level filling at which
the QHE breaks down can be determined independently by
resistance measurements of the 2DEG\cite{PG,SP}; our
explanation of the bell-shaped feature can be tested
directly in this way. This is an {\em important}
experimental test of the assumptions underlying  {\em
all} of the theories that have been proposed to date
since it should be able to determine whether the system
is fully spin polarised as has been assumed
previously\cite{TP,MG,LY,KS,KF}  or weakly spin
polarised as we have suggested here. We also predict
similar behavior at Landau level fillings below 3
(outside the range of the data in Ref. \onlinecite{Kang})
where the QHE should also break down. That the
experimental features at higher bias are asymmetric and
occur at somewhat larger bias values than the
theoretical ones can also be understood within our
framework since the QHE breaks down eventually with
increasing bias at {\em all} Landau level fillings and
does so in a {\em sample-dependent} way.~\cite{Cage}

\section{Conclusions}
\label{conc}

In conclusion, we have presented a theory that offers a
resolution of {\em all} of the puzzles posed by recent
measurements of electron transfer between the edges of
two-dimensional electron systems and have identified the
first experimental evidence of the general directed
localization regime that unifies the integer quantum
Hall effect   and one-dimensional Anderson localization.
We have also proposed a simple measurement that should
be able to settle unambiguously the important issue  of
the role that spin plays in this  system.

\acknowledgements

We thank W. Kang for correspondence and G. Narvaez
and S. Watkins for discussions. This work was
supported by the Yamada Foundation (S.N.)
and the Canadian Institute for
Advanced Research and NSERC (G.K.)



\begin{figure}
\caption{Landau levels $N$ = 0,1,2,...
become edge channels near barrier. a, Edge state energies
$E$ at zero bias vs. electron wave vector $k$ along barrier
in the simplest model of 2DEG's and barrier.
$l_B=(\hbar/eB)^{1/2}$. Left inset: 2DEG's, barrier and edge
channels. Right inset: Energy gap at edge channel crossing.
b and c, Edge channel energies for biases $V$ between the
2DEG's. Dashed lines are source and drain Fermi energies.}
\label{fig1}
\end{figure}

\begin{figure}
\caption{Electron transfer probability through barrier
$91\AA$ thick, 221 meV high\protect\cite{MG} and 18.2
microns long at $B = 6$Tesla vs. energy. Solid curve in a
(b) is for no disorder (a random potential $W$ with
correlation length $274\AA$ and $|W|<\hbar \omega_c$) at 0K.
Inset: Expanded view of narrow peak in a. Dashed curve in a
is transfer probability at 0.5K for same potential as in b;
the broad peak shows weak residual mesoscopic structure.}
\label{fig2}
\end{figure}

\begin{figure}
\caption{Comparison of theory (colored curves) and
maxima (black ellipses) of measured
conductance\protect\cite{Kang} for 2DEG's with density
$2\times 10^{11}$cm$^{-2}$. i,j indicate transfer from
(to) an edge channel derived from Landau level i(j).}
\label{fig3}
\end{figure}


\begin{references}
\bibitem{Moon}Previous experiments designed to detect
conducting channels propagating in opposite directions
along the same edge  of a 2DEG have consistently found
no evidence of this.\cite{Moon1}
\end{references}
\end{document}